%% file: hirschi_main.tex
\documentclass[multphys,vecphys]{svmult}

\usepackage{makeidx}     
\usepackage{graphicx}    
\usepackage{multicol}    

\makeindex             

\begin{document}

\title*{Pre-supernova evolution of rotating massive stars}
\author{Hirschi Raphael\inst{1}\and
Meynet Georges\inst{1}\and
Maeder Andr\'e\inst{1}\and
Goriely St\'ephane\inst{2}}
\authorrunning{Hirschi et al.} 
\institute{Observatoire de Gen\`eve,CH--1290 Sauverny, Switzerland
\texttt{raphael.hirschi@obs.unige.ch}
\and Universit\'e Libre de Bruxelles, Bruxelles, Belgique}
%
%
\maketitle
\begin{abstract}
The Geneva evolutionary 
code has been modified to study the advanced stages (Ne, O, Si burnings) of 
rotating  massive  stars.  Here we present the results of four 20\,M$_{\odot}$
  stars at solar metallicity with initial rotational
velocities, $\upsilon_{\rm{ini}}$, of 0, 100, 200 and 300\,km/s
in order to show the crucial role of rotation in stellar evolution.
 As already known, 
rotation  increases  mass  loss  and  core  masses \cite{ROTV}.  A fast rotating
20\,M$_{\odot}$ star has the   same   central   evolution  as  a
non-rotating 26\,M$_{\odot}$. Rotation also increases strongly
 net total metal  yields.  Furthermore, rotation  changes the SN type 
so that more SNIb  are  predicted  (see  \cite{ROTX}  and
\cite{PrantzosSN03}). Finally,   SN1987A--like   supernovae
progenitor colour can be explained in a single rotating star scenario.
\end{abstract}
%
%
%
\section{Computer model}
\label{sec:model}
The  computer  model used is the Geneva evolutionary code (see
\cite{ROTX}). Convective stability is determined by the Schwarzschild criterion. 
The                   overshooting                   parameter,
$\alpha_{\rm{over}}=d_{\rm{over}}/H_{\rm{P}}$ is equal to 0.1
for Hydrogen-- and Helium--burning cores 
and equal to 0 afterwards. Modifications have been made to 
study the advanced stages of the evolution of rotating massive stars. 
Dynamical   shear  has  been  included  using  $Ri_c=1/4$ 
\cite{DS02}.  Note  that  the  computer model still includes secular
shear   and  meridional  circulation.  The  structure
equations  have  been  stabilised using Sugimoto's prescription
\cite{SU70}.  Furthermore,  convection  is treated as diffusion
from the Oxygen (O) burning stage because convection is no longer instantaneous. 
 The algorithm developed for rotational mixing is used
for  this  purpose.  Finally,  the nuclear reaction network has
been  extended  and contains all the multiple-$\alpha$ elements
up to  $^{56}$Ni  except  $^8$Be.  The reaction rates are taken
from the NACRE  compilation  or Hauser-Feschbach code calculations
(ULB,  Belgium). 
\section{Evolution}
\label{sec:early}
The early evolutionary stages are presented in \cite{ROTX}. 
Here we concentrate on  
solar metallicity 20\,M$_{\odot}$  stars  and  study  the  
effect  of rotation by
examining four models with initial
rotational  velocities of 0, 100, 200 and 300\,km/s. 
Calculations have been followed until end of central O--burning   
for the $\upsilon_{\rm{ini}}=$ 100 and 200\,km/s models, 
end   of  central  Si--burning  for  the  $\upsilon_{\rm{ini}}=$
300\,km/s  model  and  end  of  first  shell Si--burning for the
non-rotating model.
\subsection{Hertzsprung--Russell (HR) diagram}
\begin{figure}
\centering
\includegraphics[height=6cm]{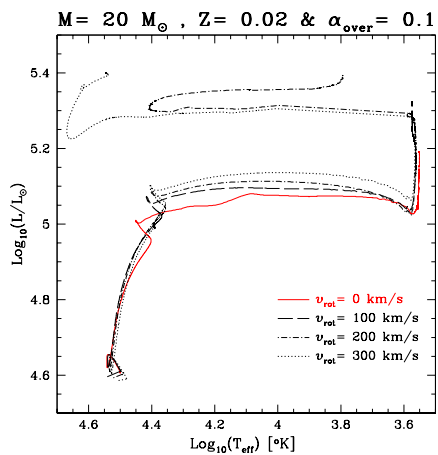}
\caption{HR diagram for the solar metallicity 20\,M$_{\odot}$ stars
 with   initial  rotational  velocities  of  0,  100,  200  and 300\,km/s.}
\label{hrcm20}       
\end{figure}
Figure  \ref{hrcm20}  shows the evolutionary tracks of the four
different   20\,M$_{\odot}$   stars  in  the  HR  diagram.  The
non-rotating model ends up as a red supergiant (RSG) like other
group models (see e.g. \cite{HL002} or \cite{LSC00}). However, the
rotating  models  show  very  interesting features. Although the
100\,km/s  model remains  a  RSG,  the  200\,km/s  model
undergoes  a  blue  loop to finish as a yellow--red supergiant
whereas  the  300\,km/s  model  ends  up as a blue supergiant (BSG). 
Thus  rotation  may  strongly affect the shock wave travel time
through  the  envelope  when  the  star explodes in a supernova
event, since this time
 is proportional to the radius of the star
(RSG radii are about hundred times BSG ones). 
Moreover,     the     behaviour     of    the    models    with
$\upsilon_{\rm{ini}}$  between 200 and 300\,km/s is reminiscent
of  the  evolution  of  the progenitor of 1987A indicating that
rotation may play a role in similar cases.
\subsection{Central evolution}
\begin{figure}
\centering
\includegraphics[height=6cm]{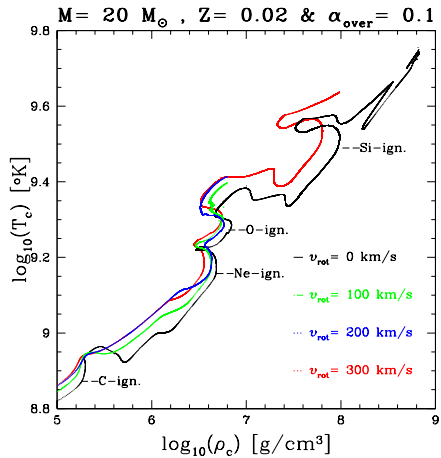}
\caption{T$_{\rm{c}}$ versus $\rho_{\rm{c}}$ diagram.}
\label{trcm20}       
\end{figure}

\begin{figure}
\centering
\includegraphics[height=5.5cm]{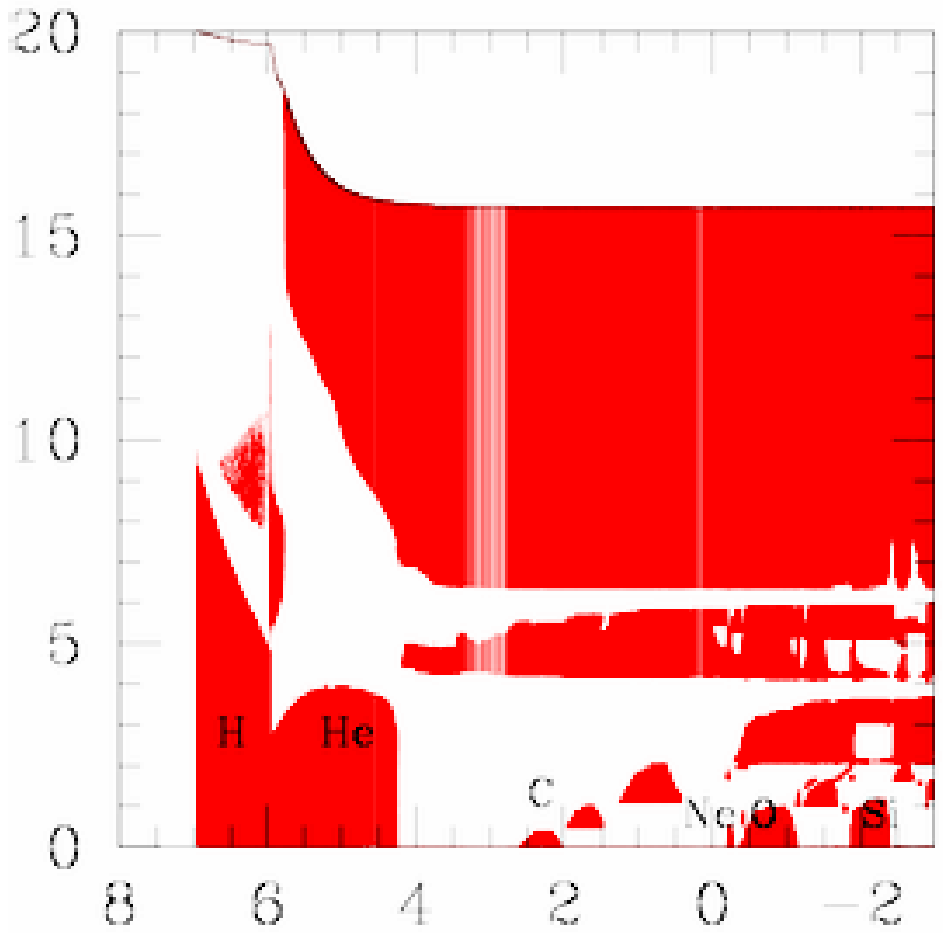}
\includegraphics[height=5.5cm]{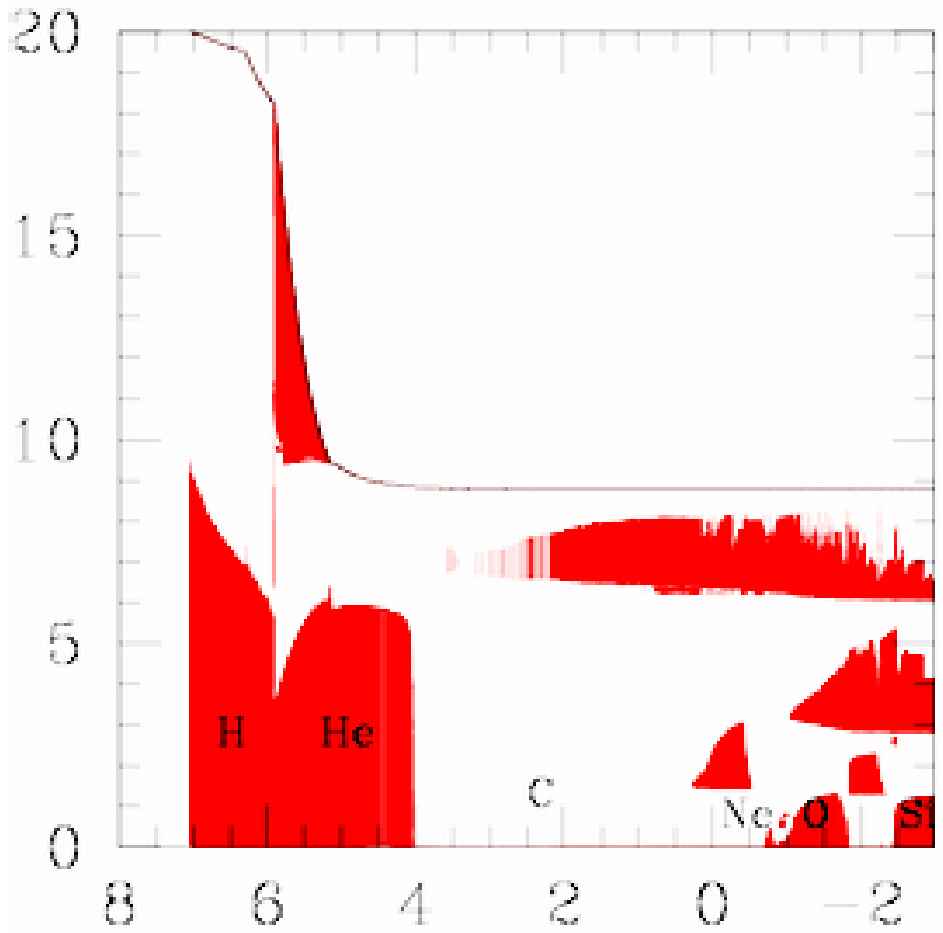}
\caption{Kippenhahn  diagrams: mass (M$_{\odot}$)  versus  
log$_{10}$($\sim$ time  left  until
core collapse) (yr). \emph{Left} non-rotating 20\,M$_{\odot}$
 model.  \emph{Right} $\upsilon_{\rm{ini}}=$ 300\,km/s 
20\,M$_{\odot}$ model. Coloured
 zones show convective zones. Letters indicate burning stages.}
\label{kips}       
\end{figure}

The  central evolution is best seen in the central temperature,
T$_{\rm{c}}$,
versus  central density, $\rho_{\rm{c}}$, diagram (Fig. \ref{trcm20}). We can see
that  rotation makes the cores slightly less degenerate (higher
T$_{\rm{c}}$ and smaller $\rho_{\rm{c}}$). This is
explained by the bigger core masses. We also see that the
``C--bump"  due  to  the convective central C--burning fades away
when rotation increases (see also Fig. \ref{kips}).
This  is  again  a consequence of more massive cores in rotating
models  which  implies  higher  neutrino loss rates and smaller
central carbon abundance at the end of He--burning phase.

\section{``Pre-SN" models}
\label{sec:late}
\subsection{Mass loss and core masses}
\begin{figure}
\centering
\includegraphics[height=6cm]{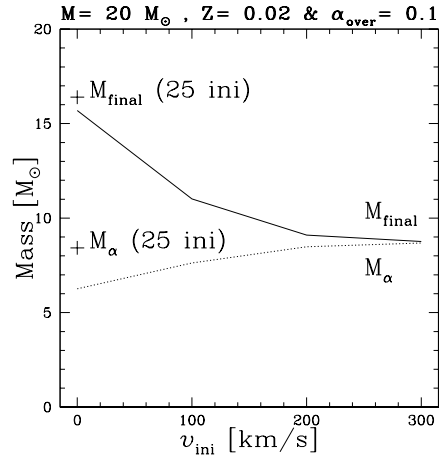}
\caption{Final total  mass,   M$_{\rm{final}}$,   and   Helium   core
mass, M$_{\alpha}$,   as   a   function  of $\upsilon_{\rm{ini}}$.
  Non-rotating 25\,M$_{\odot}$ model masses are also shown for
comparison.}
\label{m20comp}       
\end{figure}
We can see in Fig. \ref{m20comp} that both mass loss and Helium (He) core
masses,  M$_{\alpha}$,  increase  with  rotation  as already
known.  There is a saturation effect
at high rotation when the star is left with hardly any Hydrogen (H)
envelope.  As  can  be  seen  in  \cite{ROTX},  rotation
noticeably  increases the number of Wolf-Rayet stars (WR). Here
we  see  that there is a smooth transition between SN type from
IIP  $\rightarrow$  IIL  $\rightarrow$  IIb ($\rightarrow$ Ib). We
also  note  that  the $\upsilon_{\rm{ini}}=$ 
300\,km/s 20\,M$_{\odot}$ model has a
bigger He core than the non-rotating 25\,M$_{\odot}$ model
(it would correspond to the core of a non-rotating 26\,M$_{\odot}$ model).
The  Carbon--Oxygen core mass, M$_{\rm{CO}}$, increases
with rotation in a similar way as M$_{\alpha}$. The Silicium (Si) core mass
at  the  end of central O--burning only slightly increases with
rotation.
\subsection{Abundances profile and net yields}
\begin{figure}
\centering
\includegraphics[height=5.5cm]{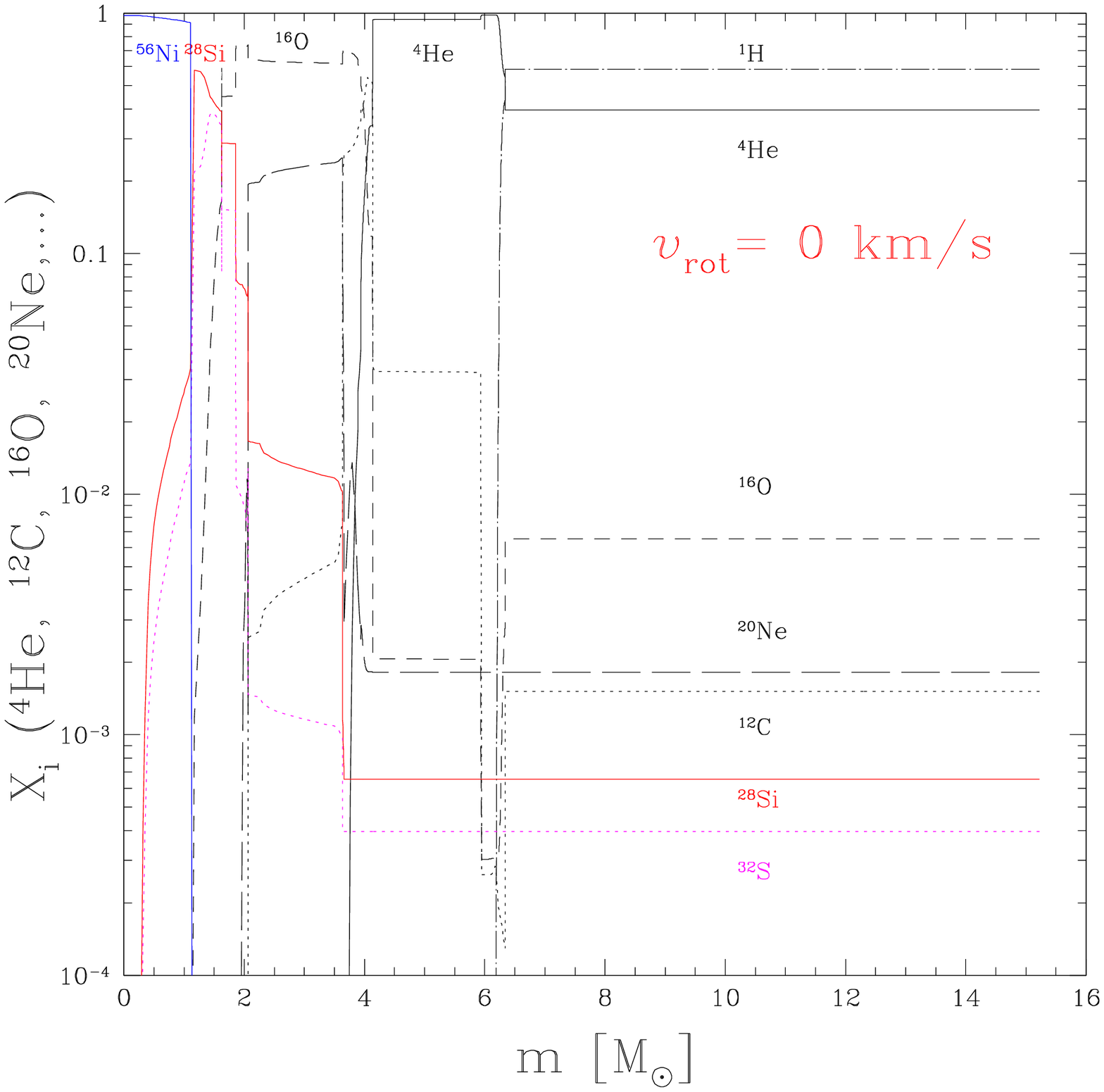}
\includegraphics[height=5.5cm]{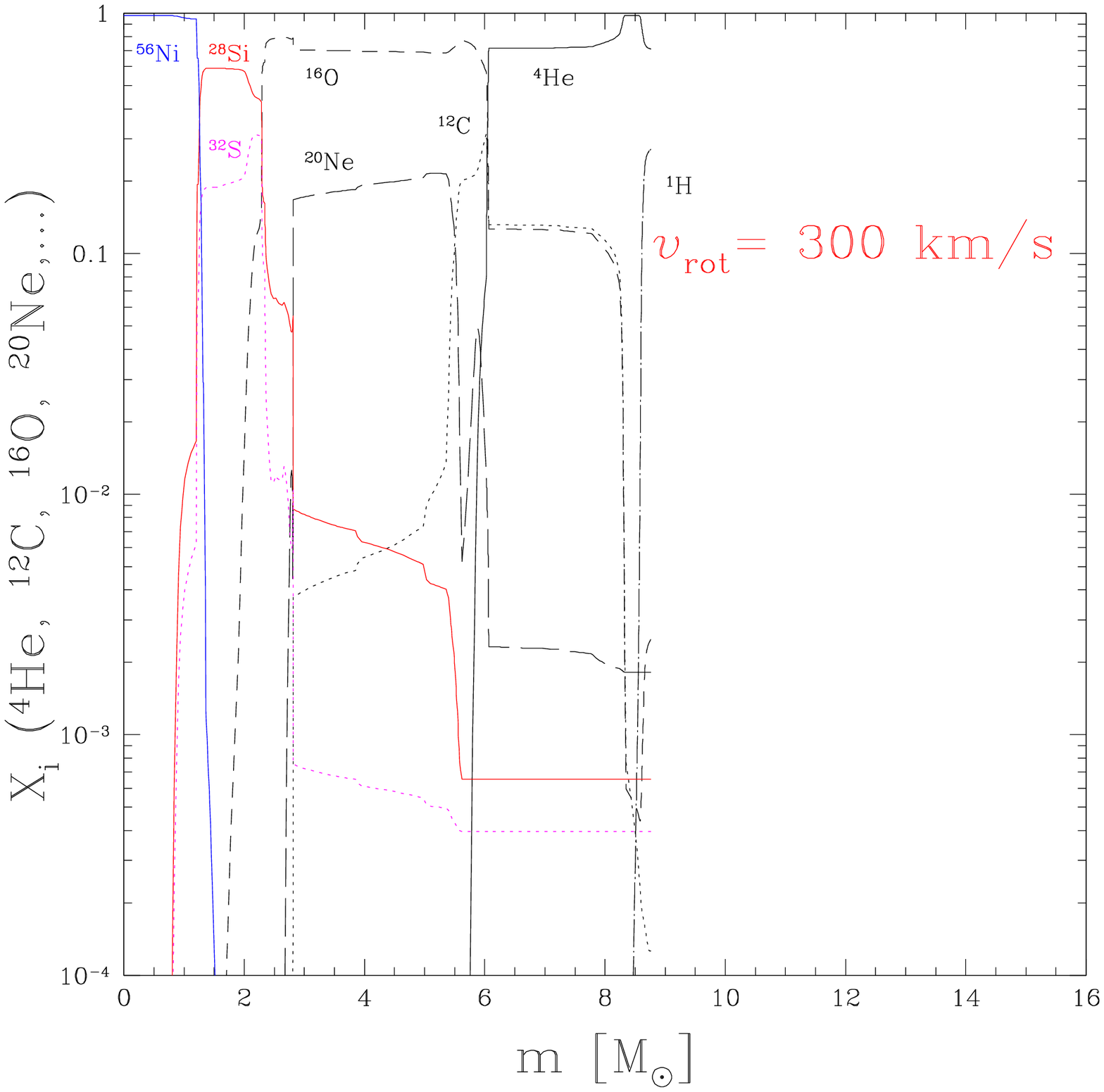}
\caption{Abundances  profile  for  main  elements at the end of
central Si--burning. \emph{Left} non-rotating 20\,M$_{\odot}$
 model.  \emph{Right} $\upsilon_{\rm{ini}}=$ 
300\,km/s 20\,M$_{\odot}$ model.}
\label{abun20}       
\end{figure}

\begin{figure}
\centering
\includegraphics[height=6cm]{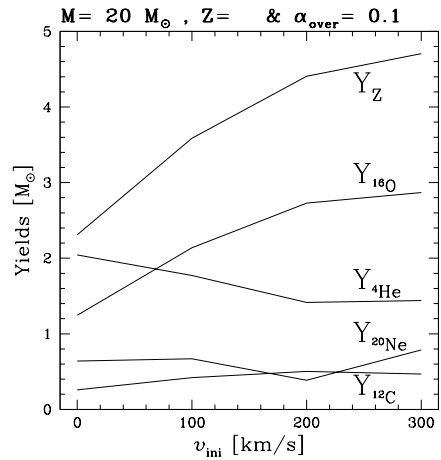}
\caption{Net  yields of
the sum of all metals, Y$_{\rm{Z}}$, and individual elements
  as a function of $\upsilon_{\rm{ini}}$.}
\label{y20comp}       
\end{figure}

In Fig. \ref{abun20} we notice the
smoother  profiles  due  to rotational mixing and also the very
small quantity of remaining H. 
The  ``pre-SN" net yields  calculated  at  this  stage  show  that
rotation  increases the total metal yield and $^{16}$O yield.
Typically, the total metal yield of the $\upsilon_{\rm{ini}}=$ 300\,km/s 
 model is twice the one of the  non-rotating model.
On the other hand, rotation decreases
H--burning products yields (notably $^4$He) as can be seen in
Fig. \ref{y20comp}.  
%
\input{hirschi_ref}

%

\printindex
\end{document}

%% file: hirschi_ref.tex
%
%